\documentclass[twocolumn,aps,superscriptaddress,preprintnumbers]{revtex4}

\usepackage{inputenc}

\usepackage{graphicx}
\usepackage{graphics}
\usepackage{dcolumn}
\usepackage{bm}
\usepackage{color}
\usepackage{textcomp}
\usepackage{amsmath}
\usepackage{amssymb}
\usepackage{braket}

\usepackage{multirow}
\usepackage{setspace}
\usepackage{ulem}

\usepackage{xcolor}

\usepackage{hyperref}
\hypersetup{
    colorlinks=true,       
    linkcolor=blue,        
    citecolor=blue,        
    filecolor=blue,        
    urlcolor=blue,         
    runcolor=blue
}

\begin{document}

\title{Model reduction for molecular diffusion in nanoporous media}

\author{Gastón A. González}
\affiliation{Department of Physics, Universidad de Santiago de Chile, Av. Victor Jara 3493, Santiago,Chile}

\author{Ruben A. Fritz}
\affiliation{Department of Physics, Universidad de Santiago de Chile, Av. Victor Jara 3493, Santiago,Chile}

\author{Yamil J. Col\'on}
\affiliation{Department of Chemical and Biomolecular Engineering, University of Notre Dame, IN, USA}

\author{Felipe Herrera}
\affiliation{Department of Physics, Universidad de Santiago de Chile, Av. Victor Jara 3493, Santiago,Chile}
\affiliation{Millennium Institute for Research in Optics, Chile}
\email{felipe.herrera.u@usach.cl}

\begin{abstract} 

Porous materials are widely used for applications in gas storage and separation. The diffusive properties of a variety of gases in porous media can be modeled using molecular dynamics simulations that can be computationally demanding depending on the pore geometry, complexity and amount of gas adsorbed. We explore a dimensionality reduction approach for estimating the self-diffusion coefficient of gases in simple pores using Langevin dynamics, such that the three-dimensional (3D) atomistic interactions that determine the diffusion properties of realistic systems can be reduced to an effective one-dimensional (1D) diffusion problem along the pore axis. We demonstrate the approach by modeling the  transport of nitrogen molecules in single-walled carbon nanotubes of different radii, showing that 1D Langevin models can be parametrized with a few single-particle 3D atomistic simulations. The reduced 1D model predicts accurate diffusion coefficients over a broad range of temperatures and gas densities. Our work paves the way for studying the diffusion process of more general porous materials as zeolites or metal-organics frameworks with effective models of reduced complexity.

\end{abstract}

\date{\today}

\maketitle

\section{Introduction}

The simulation of gas diffusion in nanoporous solid-state materials is important for applications such as gas filtering, separation and storage \cite{DPD,gordon1999molecular,daglar2018computational,krishna2005darken,COLON_2014}. The self-diffusion coefficient of a gas in a porous medium is an essential physical quantity that characterizes mass transfer and is a relevant parameter for designing industrial separation processes  \cite{liu1995accurate}, diffusion of gas mixtures  \cite{krishna2005darken}, and the selectivity of gas separation techniques \cite{koros2017materials,daglar2018computational,keil2000modeling,mace2013role,gu2018metal}. The diffusive properties of gases in porous media is ultimately related to the short and long-range interaction potentials between gas particles and between gas molecules and the condensed-phase environment \cite{tsimpanogiannis2019self}.

The growing interest in estimating the diffusive properties of target gases in porous materials reported in public databases \cite{COLON_2014} has stimulated the search for methods to accelerate large scale screening efforts based on fully-atomistic simulations, which in general are computationally demanding  \cite{sokhan2004transport,keil2000modeling,MOFs_Diffusion_2020}. Acceleration  strategies based on machine learning are promising because training sets with acceptable predictive power can be constructed with a smaller number of calculations than an exhaustive database search \cite{ML_TANG_2021,ML_ORHAN_2022}. An alternative acceleration strategy would be to develop generalizable physics-based models that are sufficiently accurate for ranking materials based on their transport properties, but at a much lower cost than atomistic simulations. 

In this context, we study the dimensionality reduction capabilities of one-dimensional (1D) Langevin dynamics for modeling gas diffusion inside carbon nanotubes at different temperatures. The predictions of the reduced model are compared to the three-dimensional (3D) molecular dynamics simulations. For concreteness, we consider the transport of molecular nitrogen in single-walled carbon nanotubes (CNT) and obtain self-diffusion coefficients with 1D Langevin dynamics for different nanotube radii, temperatures and gas densities. We show that it is possible to construct effective 1D pore potentials and model parameters that can reproduce the diffusive 3D transport behavior over a broad range of conditions. The proposed parametrization scheme could be extended to other porous materials such as zeolites and metal-organic frameworks. 

The rest of the article is organized as follows: Section \ref{Sec.Methods} describes the theoretical methodology and the settings for the atomistic molecular dynamics simulations. In Sec. \ref{Sec.Results} we discuss the results obtained for the diffusion constant of nitrogen in   carbon nanotubes, comparing the predictions of the reduced 1D Langevin model, 3D molecular dynamics simulations, and the Lifson-Jackson formula from Brownian motion theory. In Sec. \ref{Sec.Conclusions}, we suggest possible applications and generalization strategies.


\section{Methods}
\label{Sec.Methods}

\subsection{1D stochastic Langevin dynamics}

The stochastic motion of Brownian particles can be described by a Langevin equation \cite{paquet2015molecular}, which for a 1D system of $N$ particles with trajectories $z^{(\alpha)}(t)$ can be written as 
\begin{equation}\label{Eq.LE}
{{\dot{p}^{(\alpha)}}}(t)=-\dfrac{\partial V(z^N(t))}{\partial z^{(\alpha)}} - \gamma^{(\alpha)} {p}^{(\alpha)}(t) + \xi^{(\alpha)}(t) \Bigg|_{{\alpha}=1,2,.,N}
\end{equation}
where ${\alpha}$ is the particle index, $p$ is momentum, $V$ is the total potential, and $z^{(\alpha)}$ the position of the $\alpha$-th particle. The interaction of particle $\alpha$ with a large ensemble of bath particles is effectively taken into account by introducing the momentum loss (dissipation) term proportional to the damping parameter $\gamma$ and a random momentum kick given by the random process $\xi(t)$, which induces energy fluctuation. These terms together take into account the multiples collisions of the system (Brownian) particle with the reservoir \cite{DPD,paquet2015molecular}. The random momentum kick has zero bias, i.e., $\langle \xi^{(\alpha)}\rangle =0$ and its  autocorrelation function is given by
\begin{equation}\label{eq:deltacorrelation}
\langle \xi^{(\alpha)}(0)\xi^{(\beta)}(\tau)\rangle =2 \delta(\tau)\delta_{\alpha \beta}\,m^{(\alpha)}\gamma^{(\alpha)}\, k_{\rm B}T , 
\end{equation}
where $m$ is the particle mass, $k_{\rm B}$ is the Boltzmann constant, $T$ is temperature, $\delta(t)$ is the Dirac delta function and $\delta_{\alpha \beta}$ is a Kronecker delta. In other words, momentum fluctuations are Markovian in time and proportional to the thermal energy $k_{\rm B}T$. 


We solve Eq. (\ref{Eq.LE}) numerically for a system of $N$ particles using the impulsive Langevin leap-flog algorithm \cite{DPD}, which is a modification of the classical Verlet algorithm that involves an intermediate velocity correction at each time step of the form
\begin{equation}
\Delta v^{(\alpha)} = \dot{v}^{(\alpha)}h-\gamma^{(\alpha)} v^{(\alpha)}(t)h + \sqrt{2k_{\rm B} T \gamma h /m^{(\alpha)}}\xi
\end{equation}
where $v^{(\alpha)}$ and $\dot{v}^{(\alpha)}$ are the velocity and acceleration of the $\alpha$-th particle, and $h$ is the time step of the simulation. For a free Brownian particle at thermal equilibrium, the damping coefficient $\gamma$ can be obtained from the Einstein relation \cite{DPD}
\begin{equation}\label{simplediff}
D_0=\dfrac{k_{\rm B}T}{m \gamma}
\end{equation} 
where $D_0$ is the free-particle diffusion coefficient. In this work, the damping parameter $\gamma$ encodes the interaction of gas molecules with  the carbon nanotube walls.

\subsection{Diffusion from mean squared displacements}

We calculate the self-diffusion coefficient $D_s$ using the mean squared displacement (MSD) method from the simulated particle trajectories. For a trajectory composed of cartesian vectors $\vec{r}_i=(x_i,y_i,z_i)$ at times $t_i$, the MSD can be calculated as \cite{MSD} 
%
\begin{equation}\label{eq.msd}
{\rm MSD}(\tau=nh)=\dfrac{1}{M-n} \sum^{M-n}_{i=1} (\vec{r}_{i+n}-\vec{r}_{i})^2
\end{equation}
which uses all available offsets $\tau$ of a given duration $nh$ with $n$ the offset step. The advantage of this definition is that the number of such displacements is $M-n$ and therefore large for small $n$, resulting in well-averaged MSD values. MSD is related to the self-diffusion coefficient by the expression \cite{arora2006air}
\begin{equation}\label{eq_Diffusion_from_MSD}
{\rm MSD} =2aD_s\tau 
\end{equation} 
where $a$ is the system dimensionality ($a=1$ for 1D, $a=3$ for 3D). By solving Eq. (\ref{Eq.LE}) for all the particles in the system at fixed temperature and density, we calculate MSD from Eq. (\ref{eq.msd}) and obtain $D_s$ from the slope of a linear fit plot of Eq. (\ref{eq_Diffusion_from_MSD}) using the least-squares method. 

For short simulation times, particle transport is dominated by the initial condition and the absence of intermolecular interactions (ballistic regime). After equilibration is reached though multiple collisions, the linear scaling of  MSD with time is established (diffusive regime). Several methods have been proposed to analyze trajectories with coexisting transport regimes \cite{26}. In our work, the diffusive regime is established when a log-log plot MSD vs $\tau $, averaged over particles and simulation replicas has unit slope.

\begin{figure}
\centering 
\includegraphics[width=0.4\textwidth]{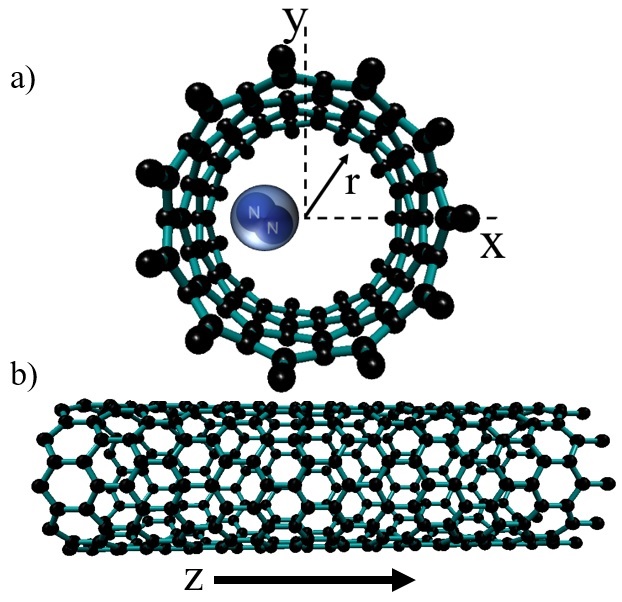}
\caption{\textbf{(a)} Radial viewpoint of the nanotube (11,0) with molecular nitrogen molecule ($N_2$) its pore volume.
\textbf{(b)} Axial viewpoint of the nanotube (11,0).} 
\label{Model}
\end{figure}

\subsection{Lifson-Jackson model for 1D diffusion}

The Lifson-Jackson formula is an analytical expression, first derived in Ref.\cite{lifson1962self}, for the diffusion coefficient of a periodic 1D potential in terms of the potential depth. The periodic nature of a pristine carbon nanotube potential along its axis allows us to use this theory directly at different temperatures. For a periodic potential $V(x)$ with period $L$, the Lifson-Jackson diffusion coefficient can be written as \cite{lifson1962self,reimann2001giant,berezhkovskii2019biased}
\begin{equation}
D'_0 (T)= \dfrac{D_0(T) L^2}{\left[\int_{-L/2}^{L/2}e^{-\frac{ V(x)}{k_{\rm B}T}}dx\right] \left[ \int_{-L/2}^{L/2}e^{\frac{ V(x)}{k_{\rm B}T}}dx \right]}
\end{equation}
where $D_0$ is the free-particle diffusion coefficient from Eq. (\ref{simplediff}). For a sinusoidal potential $V(x)=A\,{\rm sin}(ax)$ with depth $A$ and period $a/2\pi$, the integrals in the denominator can be solved analytically to give
\begin{equation}\label{Eq.CosLJ}
D'_0(T)=\dfrac{D_0(T)}{I_0^2(x)} 
\end{equation}
where $I_n(x)$ is a modified Bessel function of the first kind and $x=A/k_{\rm B}T$. Equation (\ref{Eq.CosLJ}) shows that for sinusoidal potentials, self-diffusion is determined by the ratio between the depth of the potential and the thermal energy, independent of the lattice period. $D'_0$ reduces to the free-particle limit at high-temperatures, and asymptotically vanishes at low temperatures, as inferred from the asymptotic forms $I_0(x\rightarrow 0)\sim 1$  and $I_0(x\rightarrow \infty)\sim \infty$.

\begin{figure}[t]
\centering 
\includegraphics[width=0.5\textwidth]{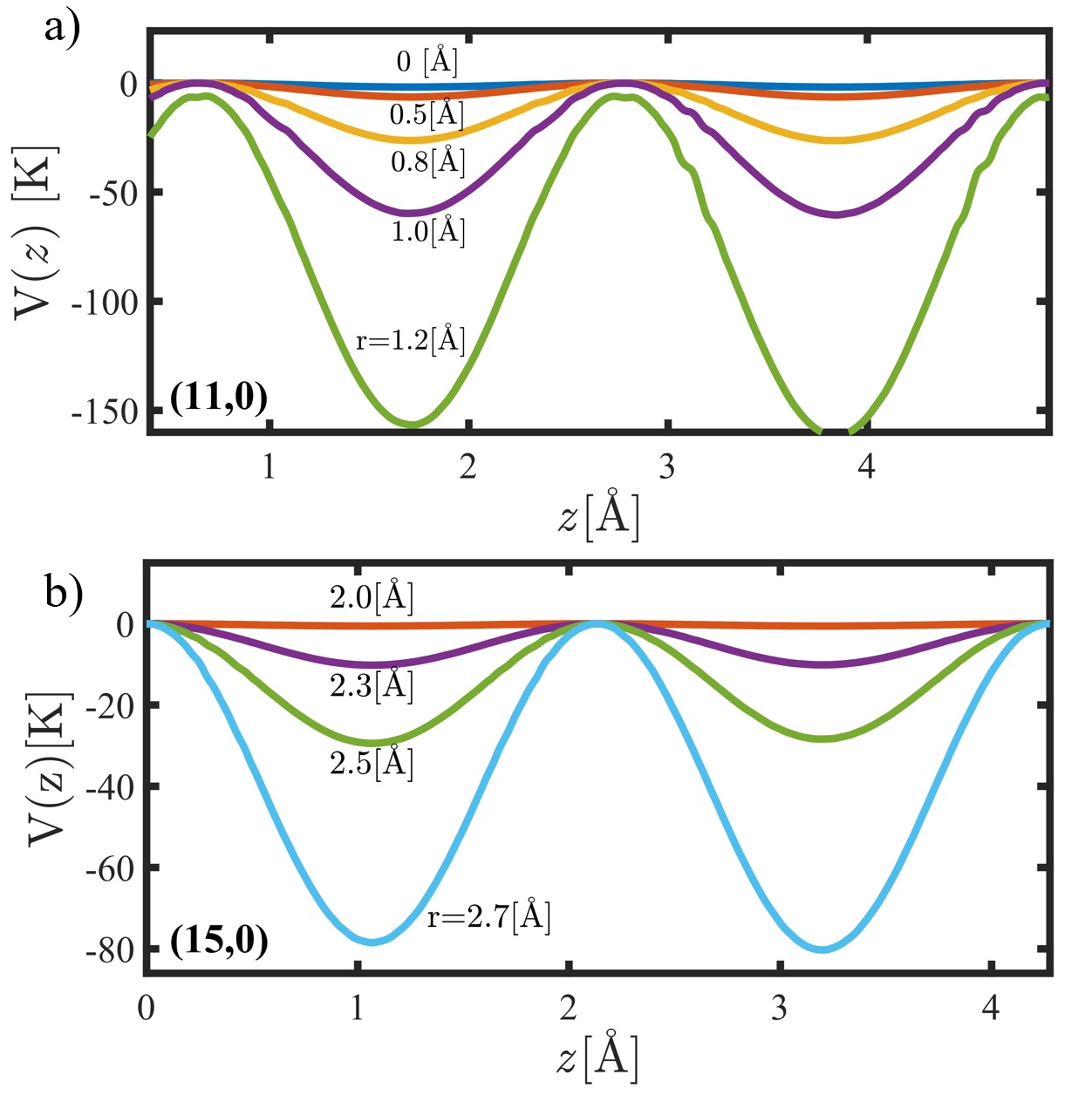}
\caption{ \textbf{(a)} Axial potentials at different radii from the center of the nanotube (11,0).
\textbf{(b)} Axial potentials for (15,0). Curves are labelled by the values of the radial coordinates         }

\label{Fig.axial}
\end{figure}

\subsection{Molecular nitrogen in carbon nanotubes}

The self-diffusion of molecular nitrogen inside a carbon nanotube was computed over a broad range of temperatures in the range $1-10^3$ K, and a range of gas densities spanning from the single-molecule limit to pore saturation. We perform calculations using zig-zag carbon nanotubes (11,0) and (15,0), with radii 4.309 \r{A} and 5.876 \r{A}, respectively. The nanotube coordinates were obtained with a modeler software \cite{NTM}, for a tube length of 426.3 \r{A}. Figure \ref{Model} shows representative radial and axial views of the nanotubes used in this work. For MD and Langevin dynamics simulations, the thermal motion of carbon atoms in the nanotubes were ignored, which does not introduce significant errors in the evaluation of gas diffusion constants. We set periodic boundary conditions, random initial locations of the gas particles, and a thermalization time of 0.5 ns in all simulations. 

We model the interaction between nitrogen molecules and between nitrogen and carbon atoms in the nanotube with a Lennard-Jones potential
\begin{equation}
V(R)=4\epsilon \left[ \left( \dfrac{\sigma}{R}   \right)^{12}-\left( \dfrac{\sigma}{R}   \right)^6  \right]
\end{equation}
where $R$ is the interparticle distance. The potential parameters for each particle pair in the problem are listed in Table \ref{A.Tab.LJ}.
\begin{table} [t]
\begin{center}
\begin{tabular}{| c | c | c | c | c |}
\hline
		 			& $N-N$ & $C-N$ & $N_2-N_2$ & $C-N_2$\\ \hline
$\sigma$ (\r{A}) 		& 3.32 & 3.36 &   3.63      & 3.52   \\
$\epsilon /k_B (K)$ & 36.4 & 33.4 &   104.5     & 56.2   \\ \hline
\end{tabular}

\end{center}
\caption{Lennard-Jones parameters. $N-N$ and $C-N$ used in LAMMPS \cite{arora2006air}. $N_2-N_2$ and $C-N_2$ used in 1D Langevin.} 
\label{A.Tab.LJ}
\end{table}



MD simulations are implemented in LAMMPS  \cite{lammps}. To compute 3D MSD trajectories, we adopt a non-vibrating diatomic molecule model  for nitrogen, with three-dimensional rotational and translational motion inside the CNT.  We use a time step $h=1$ fs in the canonical ensemble. Each replica corresponds to a total simulation time of 5 ns. Depending on the depth of the effective axial potential experienced by a molecule in the nanotube, at very low gas densities (single particle) there is a temperature in which nitrogen molecules behave as quasi-free Brownian particles, as seen from the linear scaling of the diffusion coefficient with temperature. In this regime, we assume that the Einstein relation in Eq. (\ref{simplediff}) holds and extract the effective damping parameter  $\gamma$ from a linear fit. For higher gas densities, nitrogen molecules are added inside the CNT with random locations and orientations. For MD simulations we define the filling ratio $\eta = \rho/\rho_0$ to quantify nitrogen density $\rho$ relative to the tabulated density of bulk liquid nitrogen $\rho_0$ at the simulation temperature.


The stochastic 1D simulations were implemented in Matlab with the impulsive Langevin leap-flog algorithm \cite{DPD}, as mentioned previously. As input for the simulation we constructed axial potentials $V(z)$ that capture the interaction of nitrogen molecules with the CNT walls along the transport direction. In Figure \ref{Fig.axial} we show effective axial potentials constructed for nanotubes (11,0) and (15,0) at different radial distances from the nanotube center. The potentials are periodic with a lattice constant of about $2.1$\r{A}, which correlates with the equilibrium carbon-carbon distance in the nanotubes. At the center of the nanotubes, the depth of the axial potential becomes negligible, and is higher near the walls.

\begin{figure}[t]
\centering 
\includegraphics[width=0.5\textwidth]{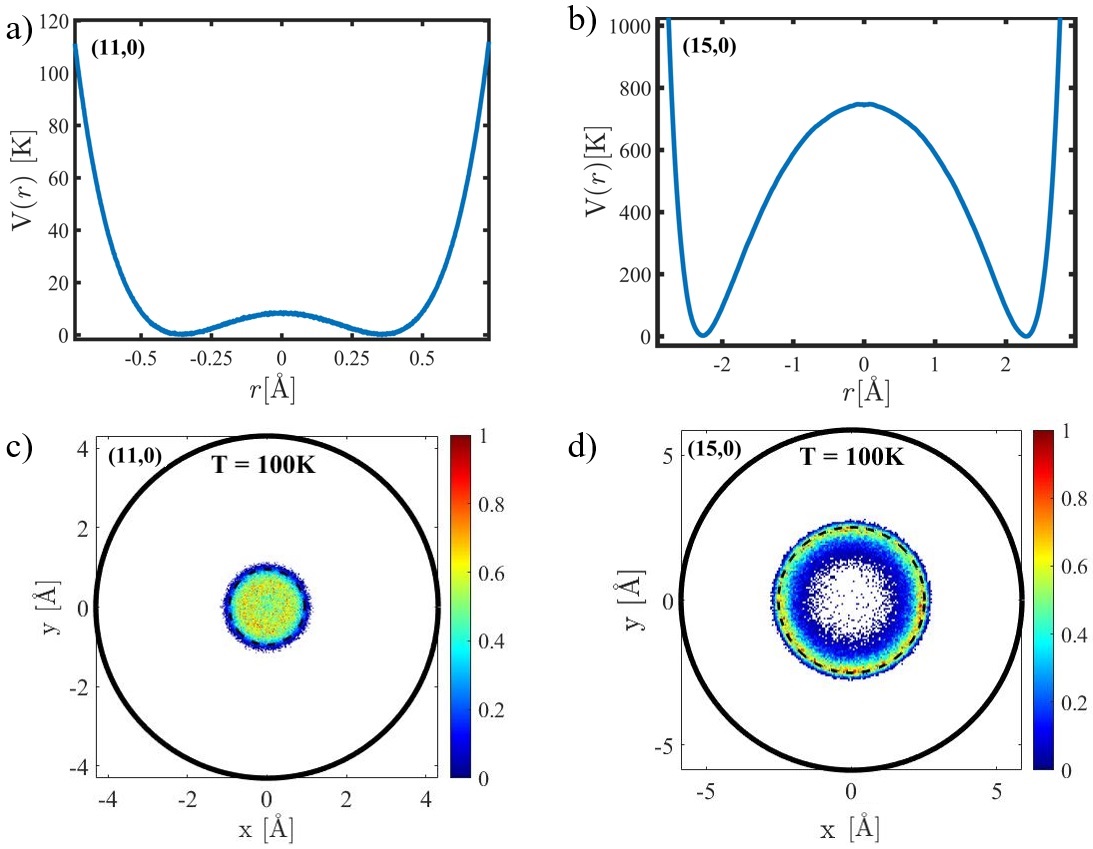}
\caption{\textbf{(a)} Effective radial potential of SWCNT(11,0) with a potential barrier in the center of approximately 10K.
\textbf{(b)} Effective radial potential of SWCNT(15,0) with a potential barrier in the center of approximately 700K.
\textbf{(c)} 2D Histogram of the nitrogen positions inside nanotube (11,0)
\textbf{(d)} 2D Histogram of the nitrogen positions inside nanotube (15,0), at 100K and bar shows normalized number of counts, solid black circle represents the centers of the CNT atoms, black dotted line represents repulsive regime of CNT atoms.   } 
\label{Fig.RPotential}
\end{figure}

In Figure \ref{Fig.RPotential} we show representative radial potentials for the nanotubes (11,0) and (15,0). The potentials feature a repulsive wall near the nanotube radius and radial barrier at the center that separates two potential minima with azimuthal symmetry. The central barrier is about 10 K high for (11,0), and 700 K for (15,0). In Figs. \ref{Fig.RPotential}c and \ref{Fig.RPotential}d we show the histograms of the radial positions that  nitrogen molecules explore at 100K, as obtained from 3D MD trajectories. While for (11,0), the nitrogen molecules tend to move near the center of the nanotube, for (15,0), the nitrogen molecules tend to move around the minimum of the radial potential, which has ring shape along the azimuthal coordinates. Practically no trajectories explore the nanotube center in this case. 

For projecting the nitrogen molecule degrees of freedom to 1D axial motion, we replace the rotating diatomic nitrogen by a spherical mass at the position of the molecular center of mass, as illustrated in Fig. \ref{Model}a. However, the Lennard-Jones parameters in Table \ref{A.Tab.LJ} do take into account the orientational dependence of the interaction potential between two nitrogen molecules and between nitrogen diatomic and carbon atoms through a thermal averaging procedure described in the Supplementary Material (SM). The stochastic MSD trajectories were obtained with a damping parameter $\gamma$  calibrated from a dilute nanotube MD simulation, as previously described. The 1D simulation time step is $h=30$ fs. The total simulation time is 6.5 ns. To define a 1D filling ratio, we assume the nanotube is saturated ($\eta \approx 1$) when the number of nitrogen molecules in the simulation is equal to the ratio between the van der Waals diameter of molecular nitrogen and the length of the simulation box.

%

\begin{figure*}[t]
\centering 
\includegraphics[width=14.0cm]{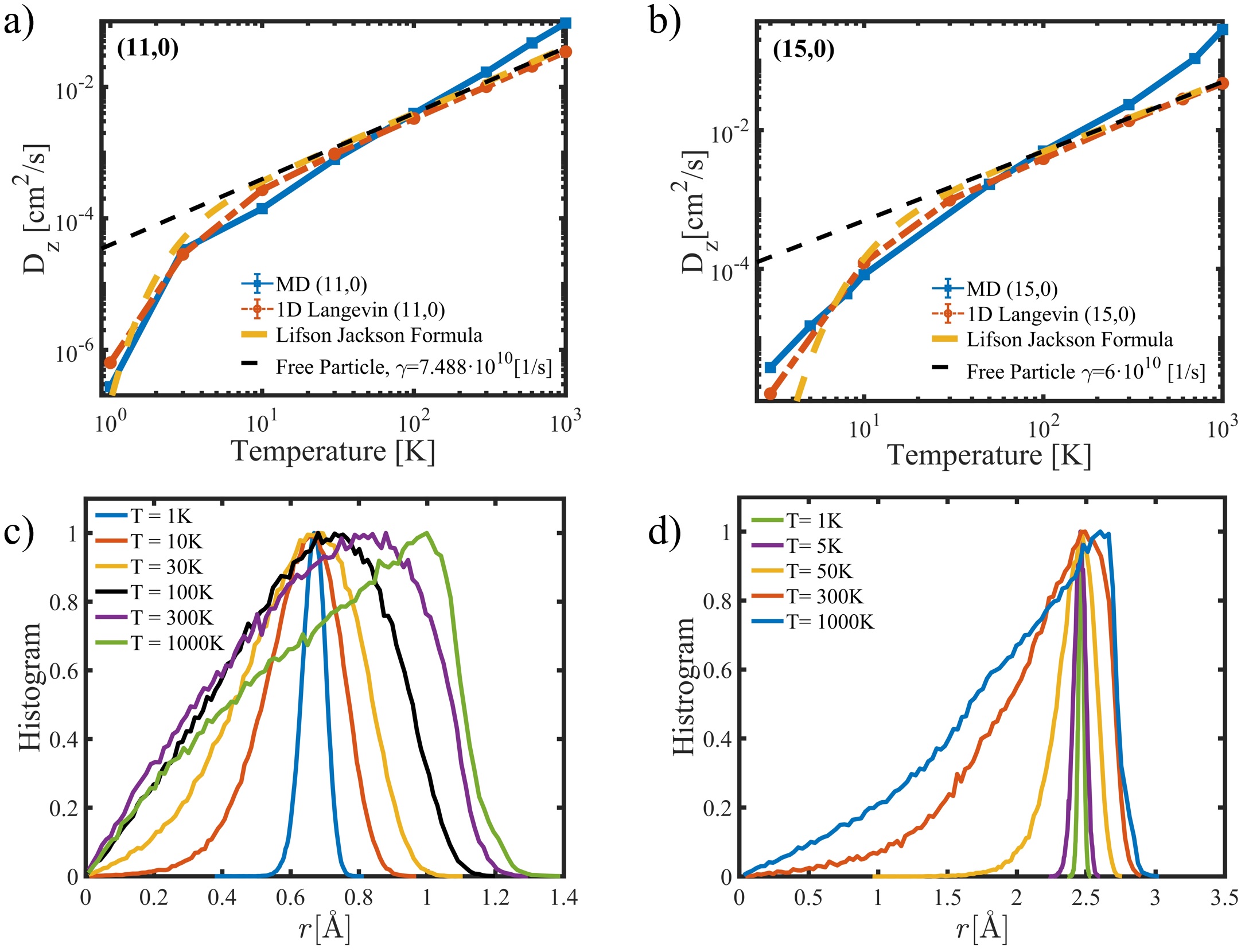}
\caption{ \textbf{(a)} 
Single Particle  Diffusion vs Temperature of CNT (11,0). 
\textbf{(b)} 
Single Particle  Diffusion vs Temperature of CNT and (15,0). Blue: LAMMPS Diffusion; Red: Langevin Diffusion; Yellow: Lifson Jackson Formula; Black: Free Brownian particle with $\gamma = 7.488 \times 10^{10}$ and $\gamma = 6.0 \times 10^{10}$ respectively.
\textbf{(c)} 
Normalized histogram of the center of mass of each nitrogen molecule inside nanotube (11,0).
\textbf{(d)} 
Normalized histogram of the center of mass of each nitrogen molecule inside nanotube (15,0). Curves are labelled by the value of the temperature.
}
\label{Fig.4}
\end{figure*}


\section{RESULTS AND DISCUSSION}
\label{Sec.Results}


\begin{figure*}
\centering 
\includegraphics[width=0.9\textwidth]{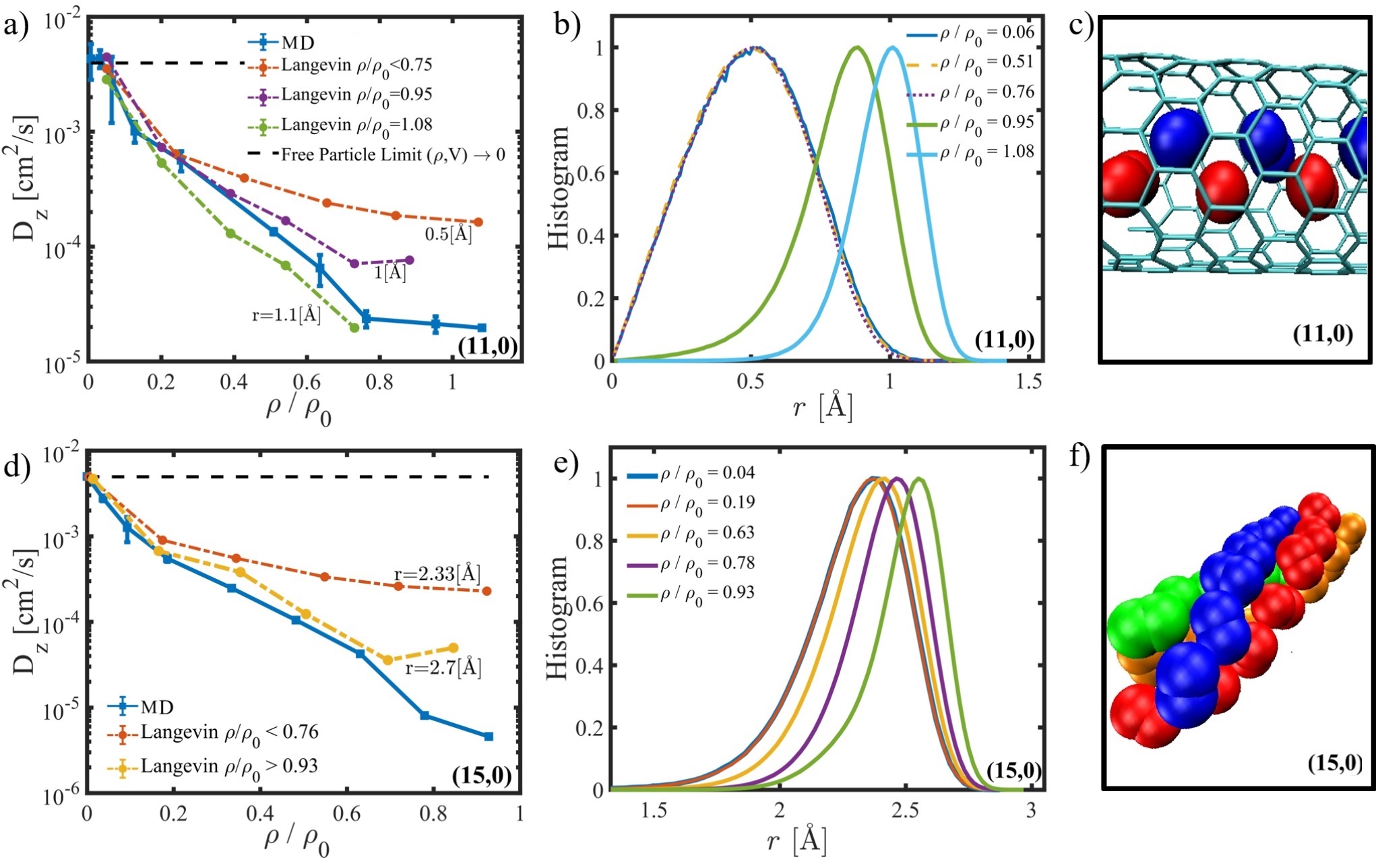}
\caption{\textbf{(a)} Diffusion vs density (11,0) at 100K. Blue: MD. Red, Green and purple: Langevin diffusion with different depths of axial potentials as a function of the radial distance from the center of the nanotube . Black: free particle limit according to Eq. (\ref{simplediff}) with $\gamma = 7.488\times 10^{10}$. 
\textbf{(b)} Histogram of the radial positions of nitrogen molecules at different densities in MD of CNT (11,0).
\textbf{(c)} Graphical modeling of the nitrogen molecules inside the (11,0) nanotube at 1.1 density calculated, made with VMD \cite{HUMP96}
\textbf{d)}  Diffusion vs density (15,0) at 100K. Blue: MD. Red, Green and purple: Langevin diffusion with different depths of axial potentials as a function of the radial distance from the center of the nanotube . Black: free particle limit according to Eq. (\ref{simplediff}) with $\gamma = 6.0\times 10^{10}$.
\textbf{(e)} Histogram of the radial positions of nitrogen molecules at different densities in MD of CNT (15,0). 
\textbf{(f)} Graphical modeling of the nitrogen molecules inside the (15,0) nanotube at 0.93 density calculated, made with VMD \cite{HUMP96} } 
\label{Multi}
\end{figure*}
 

In Fig. \ref{Fig.4} we show the self-diffusion coefficient for a single nitrogen molecule in carbon nanotubes (11,0) and (15,0), as a function of  temperature. We compare the results obtained from 3D MD simulations, 1D Langevin simulations and the Lifson-Jackson formula. The latter is evaluated with the same axial potential used in the Langevin calculations. The damping parameter obtained via linear fit from the MD diffusion coefficient at 100 K was $\gamma=7.5 \times 10^{10}\, {\rm s}^{-1}$ for (11,0), and $\gamma=6.0 \times  10^{10}\, {\rm s}^{-1}$ for (15,0).   

Below $\sim 3$ K there is essentially no diffusion in the nanotubes, because the thermal energy is lower than the corresponding axial potential depths (see Fig. \ref{Fig.axial}), so particles become trapped in the axial potential. At higher temperatures, the MSD trajectory analysis gives a diffusive regime with a log-log slope of $1\pm0.15$ \cite{arora2006air}, from which we obtain converged diffusion coefficients. 

As the temperature increases, all methods capture a crossover between particle trapping and diffusion around $10$ K, beyond which the diffusion constant scales linearly with temperature, as expected from Eq. (\ref{simplediff}) for quasi-free Brownian motion. The self-diffusion coefficient for 1D langevin and the Lifson-Jackson formula coincide in the entire range of temperatures studied. However, while the orders of magnitude are the same, the 3D diffusion coefficients obtained with MD are consistently greater. {The similarity in self-diffusion coefficients of the 1D Langevin and Lifson Jackson formulas can be explained by the fact that both methods use the same one-dimensional axial potential as input data. In addition, at high temperatures they must converge to the value of $D_0$}. The discrepancy between 3D and 1D results grows with temperature, as Fig.\ref{Fig.4}b illustrates more clearly for the wider (15,0) nanotube. 

To understand this discrepancy at high temperature, in Figs. \ref{Fig.4}c and \ref{Fig.4}d we show the histograms of the radial positions of the centers of mass explored by nitrogen molecules  at different temperatures for the nanotubes (11,0) and (15,0), obtained from 3D MD trajectories. At low temperatures ($T\sim 10$ K), molecules are mostly trapped at the minima of their corresponding radial potentials. At higher temperatures  ($T\sim 100$ K), particles have more energy to explore a larger fraction of the nanotube pore volume, broadening the radial distribution and displacing the most-probable radius towards the nanotube walls. This effective increase in the configuration space involved in axial transport cannot be captured by the effective 1D Langevin model, without redefining $\gamma$. However, the agreement is excellent between the dilute 1D Langevin with a single value of $\gamma$ and the 3D MD simulations, over a broad range of temperatures.

In Figs. \ref{Multi}a and \ref{Multi}b we plot the nitrogen diffusion coefficient as a function of the gas filling ratio $\eta=\rho/\rho_0$ for (11,0) and (15,0) nanotubes, respectively. We compare the results obtained from 3D MD and 1D Langevin simulations at 100 K. For the 1D calculations, we approximately capture the density-dependence of the effective axial molecule-nanotube axial potential by evaluating the nitrogen-nanotube axial potential at the peak of the radial distribution of MD trajectories shown in Fig. \ref{Multi}b for (11,0) and Fig. \ref{Multi}e for (15,0). For increasing molecular densities, close to saturation ($\eta>0.90$). Radial density of trajectories peak closer to the pore walls. In general, both 3D and 1D simulations give diffusion coefficients that decrease monotonically with the pore occupation for the two nanotube radii considered. 

Depending on the radial position used to estimate the depth of the effective axial potential $V(z)$, the Langevin calculations can approximate the atomistic 3D results reasonably well over the entire range of densities up to the saturation regime ($\eta \sim 1$).  For axial potential depths below 30 K ($r<0.8$ \r{A}), Fig. \ref{Multi}a shows that the agreement between the 1D and 3D curves is excellent up to $\eta\approx 0.4$ for (11,0) nanotubes.  At these low gas concentrations, nitrogen molecules move preferably near the center of the nanotube (see peak at $r=0.5$ \r{A} in Fig. \ref{Multi}b).

At higher densities ($\eta\sim 0.8-1.0$), there is a sudden shift in radial density towards the walls of the (11,0)  nanotube. This shift is due to emergence of stacked configurations between nitrogen molecules, as shown in Fig. \ref{Multi}c. At higher densities, the Langevin simulations consistently overestimate the diffusion coefficient relative to atomistic MD, although both 1D and 3D continue to have similar qualitative behavior, reaching asymptotic saturation for $\eta\sim 1$. By evaluating the axial potential closer to the peak of the radial trajectory distribution at the corresponding density (see Fig. \ref{Multi}b), the Langevin results can be made to agree with the MD simulations over a wider range of densities with the same low-density value of $\gamma$. 

For the wider (15,0) nanotube, we find similar trends when comparing 1D and 3D diffusion coefficients in Fig. \ref{Multi}d. Again the agreement between MD and Langevin simulations can be improved by sampling the axial potential closer to the peak of the radial trajectory distribution at a given gas density (Fig. \ref{Multi}e). The main qualitative difference between (11,0) and (15,0) nanotubes occurs near saturation, as the larger pore volume of (15,0) allows for more intricate stacking configurations of the nitrogen molecules, which are more difficult to capture with 1D effective models than the small-pore saturation behavior of (11,0), for example. Fig. \ref{Multi}f shows a representative quadruple ``helix" configuration that nitrogen molecules adopt at high filling ratios in the (15,0) nanotube ($\eta= 0.9$). These helical structures have been reported in carbon nanotubes for nitrogen \cite{arora2006air} and water  \cite{helicalwater}.


\section{Conclusions}
\label{Sec.Conclusions}

In this work we developed an effective one-dimensional Langevin equation model for the diffusive transport for dilute and dense molecular gases inside carbon nanotubes, as a function of tube radius and temperature. By parametrizing the Langevin model  using atomistic molecular dynamics simulations over a limited range of densities and temperatures, we find that the reduced stochastic approach can accurately extrapolate the behavior of the diffusion coefficient over a broader range of temperatures and nanotube filling ratios. For higher densities closer to saturation, we show that the effective potential that drives the Langevin dynamics along the nanotube axis can be adjusted to account for the interaction between gas particles over transverse degrees of freedom, and propose criteria to obtain effective Langevin potentials and damping parameters using nitrogen transport in carbon nanotubes as an example. We envision future extensions of the proposed dimensionality reduction methodology to study diffusive transport of gases and liquids in complex nanoporous media such as metal-organic frameworks, zeolites and structured electrodes, which could facilitate the large-scale screening of materials for applications in energy, catalysis, and gas separation.


\section{acknowledgements}

RAF is supported by DICYT-USACH grant POSTDOC USA1956 \_ DICYT and ANID Fondecyt Postdoctoral 3220857. FH and GG are supported by ANID through Fondecyt Regular 1181743 and Millennium Science Initiative Program ICN17-012. YJC thanks the University of Notre Dame for financial support through start-up funds.

\bibliographystyle{unsrt}
\bibliography{LangevinDPD}


\appendix
\begin{widetext}
\newpage
\section{Effective Potential}
\label{Apx.Potential}

We want to find the effective potential between the diatomic molecules of $N_2$ and $C-N_2$. If we simulate a large set of potentials of $N_2-N_2$ considering all possible configurations or orientations with equal probability, we obtain a wide range of potential values, as is shown in Figure\ref{FIG.APX}a.
Linearly adding the interactions between each pair of molecules ($r_{13},r_{14},r_{23},r_{24}$) for random orientations of each molecule (\ref{FIG.APX}b) and then averaging with a Boltzman weight, it is possible to find an effective total potential at a certain temperature.

\begin{equation}
V(T,d)=\frac{\sum_i e^{\frac{-min(V_i(d))}{k_BT}} \cdot V_i(d)}{\sum_i e^{\frac{-min( V_i(d))}{k_BT}}}
\label{Eq.BoltzMean}
\end{equation}

This effect is easily explained if we consider the effect of the spatial orientations of the $N_2$ molecules. At low temperatures, the possible spatial orientations experienced by the $N_2$ molecule are "limited". They are arranged in such a way as to minimize energy. The opposite is the case at high temperatures that experience almost equally all possible spatial orientations, including (for example) a system of two interacting $N_2$ molecules arranged collinearly in space (system where energy is maximized).
Replicating the previous calculation, the interaction potential between $C-N_2$ can be determined, we can calculate the potential as a function of the radius of the interior of the nanotube.
Finally, by finding the minima of the potentials and their intersection with zero, we can determine their potential analog of Lennard Jones, the parameters obtained are found in Table. \ref{A.Tab.LJ}.\\
\begin{figure}
\centering 
\includegraphics[width=0.9\textwidth]{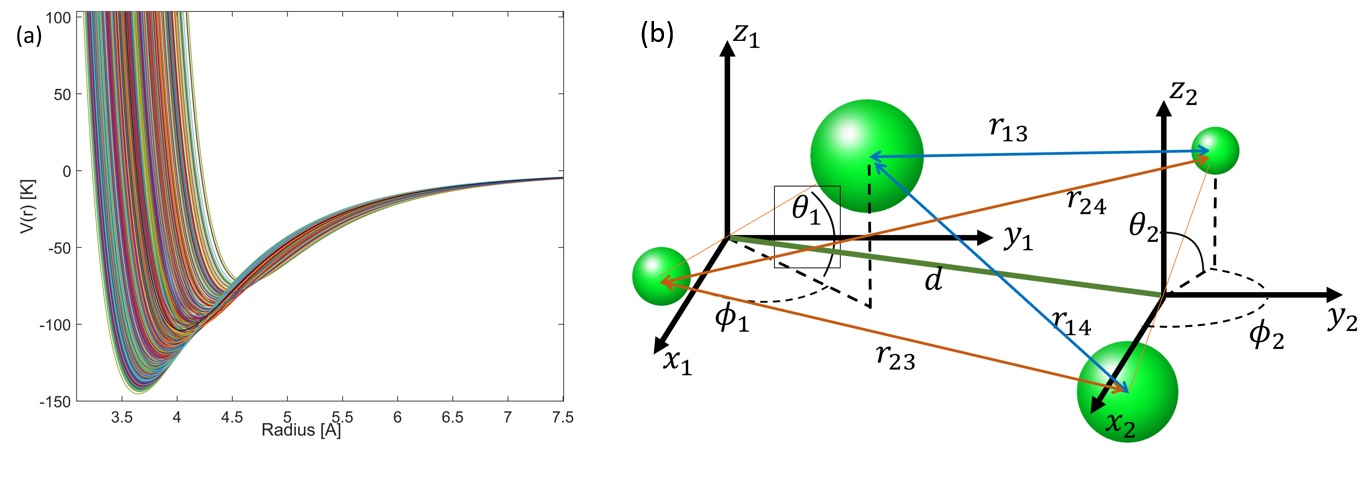}
\caption{\textbf{(a)} potentials of $N_2-N_2$ considering many possible configurations or orientations. \textbf{(b)} Scheme of spatial positions between 2 nitrogen molecules  } 
\label{FIG.APX}
\end{figure}
\end{widetext}

\end{document}